\documentclass[amsmath,amssymb,aps,prl,reprint,groupedaddress,showpacs]{revtex4-1}

\usepackage{multirow}

\newcommand{\Proof}{\noindent\textbf{Proof.}\quad}
\newcommand{\qed}{\hfill$\Box$}
\newcommand{\ket}[1]{\left\vert #1 \right\rangle}
\newcommand{\bra}[1]{\left\langle #1 \right\vert}

\newtheorem{theorem*}{Theorem}
\newtheorem{theorem}{Theorem}
\newtheorem{lemma}[theorem]{Lemma}
\newtheorem{corollary}[theorem]{Corollary}
\newtheorem{proposition}[theorem]{Proposition}

\begin{document}

\title{Quantum error correction via less noisy qubits}

\author{Yuichiro Fujiwara}
\email[]{yuichiro.fujiwara@caltech.edu}
\affiliation{Division of Physics, Mathematics and Astronomy, California Institute of Technology, MC 253-37, Pasadena, California 91125, USA}

\date{\today}

\begin{abstract}
Known quantum error correction schemes are typically able to take advantage of only a limited class of classical error-correcting codes.
Entanglement-assisted quantum error correction is a partial solution which
made it possible to exploit any classical linear codes over the binary or quaternary finite field.
However, the known entanglement-assisted scheme requires noiseless qubits
that help correct quantum errors on noisy qubits, which can be too severe an assumption.
We prove that a more relaxed and realistic assumption is sufficient
by presenting encoding and decoding operations assisted by qubits on which quantum errors of one particular kind may occur.
As in entanglement assistance, our scheme can import any binary or quaternary linear codes.
If the auxiliary qubits are noiseless,
our codes become entanglement-assisted codes,
and saturate the quantum Singleton bound
when the underlying classical codes are maximum distance separable.
\end{abstract}


\maketitle

Quantum error correction is one of the most important building blocks for reliable large-scale quantum computation and quantum communication.
Since the discovery of the fact that the effect of decoherence can be reversed \cite{Shor,Steane},
the theory of quantum error correction has made rapid and remarkable progress
including experimental realizations \cite{CPMKLZHS,KLMN,Cetal,BVFC,Schindleretal,MBRL,RDNSFGS,TopoQEC,ZLS,ZGZL}.

On the theory side, the most extensively studied class of quantum error-correcting codes is \textit{stabilizer codes} \cite{Gdissertation}.
They may be regarded as a quantum analogue of the fundamental error-correcting codes in classical coding theory, namely the \textit{linear codes} \cite{HP}.
In fact, stabilizer codes can be constructed from linear codes.

The striking difference between classical linear codes and the quantum counterpart is that
quantum error correction imposes a severe constraint on the possible structures of a code.
This forbids the use of the vast majority of linear codes as stabilizer codes.

The discovery of the \textit{entanglement-assisted stabilizer formalism} is a relatively recent development that proved that
the severe structural constraint can be circumvented if maximally entangled states are preshared between the information source and sink \cite{BDH}.
In other words, one may import any linear codes over the finite field of order two or four
if some qubits are transmitted through a noiseless channel.
This led to research on how to import excellent classical codes
while consuming only a tiny number of noiseless qubits \cite{HDB,WB,HBD,DDJCY,Djordjevic4,FCVBT,WB3,WB4,HYH2,LB,FV}.

While in principle the entanglement-assisted stabilizer formalism allows for directly exploiting classical coding theory,
it is not a panacea for quantum error correction; noiseless qubits are extremely difficult to provide, however few they may be.
This disadvantage is more pronounced in computing scenarios such as protecting quantum memory, where
the information source and sink are separated in the time domain.

Since asking for completely noiseless qubits is very demanding,
one may ask if it is enough to provide some qubits which are less noisy in some way:
Can we freely import linear codes if we have access to a less noisy, more realistic auxiliary quantum channel?
This Letter answers this question in the affirmative.

We would like to protect quantum information from general quantum errors.
Roughly speaking, the most general quantum error is described
by a linear combination of a bit error, phase error, and both at the same time \cite{MikeandIke}.
With the process called \textit{discretization}, it is enough to be able to reverse the effect of the Pauli operators $X$ and $Z$,
which correspond to a bit error and a phase error respectively.
We employ this discretization technique.
Hence, our focus is on errors $X$ and $Z$.

An important point to note is that not every kind of quantum error is equally easy to suppress through technical developments on hardware.
For instance, phase errors are expected to be far more likely than bit errors in actual quantum devices
(see, for example, \cite{IM} and references therein).
Hence, if we are allowed to assume that some physical qubits are more reliable than others,
it is more realistic to assume that the less noisy qubits may not be completely noiseless but only suffer from the dominating type of quantum error.

The primary purpose of the present Letter is to prove that this assumption is enough to exploit linear codes.
We present encoding and decoding operations that turn any classical linear code over the finite field of order two or four
into a quantum error-correcting code under the assumption that
there is a fixed set of qubits on which only the phase error operator $Z$ acts during information transmission.
More specifically, we will prove the following theorem and its extension to the binary case:
\begin{theorem}\label{onlyZ}
If there exists a linear $[n,k,d]_4$ code over $\mathbb{F}_4$,
then there exit unitary operations that encode $k$ logical qubits into $2n-k$ physical qubits and correct up to $\lfloor\frac{d-1}{2}\rfloor$ quantum errors
under the assumption that a fixed set of $2(n-k)$ physical qubits may experience phase errors but no bit errors.
\end{theorem}
Our method can also be modified so that the auxiliary qubits only suffer from the bit error operator $X$ in a straightforward manner.
If we assume that the less noisy qubits are free from any kind of quantum error,
our scheme provides entanglement-assisted quantum error-correcting codes.
As in the standard method for realizing entanglement-assisted codes,
these codes saturate the upper bound on the error correction performance called the quantum Singleton bound \cite{KL}
if maximum distance separable (MDS) codes \cite{HP} are employed as the underlying classical codes for quantum error correction.

\textit{Preliminaries.---}
We first introduce a variant of syndrome decoding for classical linear codes.
For the basics of classical and quantum error correction, we refer the reader to \cite{HP,MikeandIke}.

Take the finite field $\mathbb{F}_4 = \{0,1,\omega,\omega^2=\omega+1\}$ of order four and its prime subfield $\mathbb{F}_2 = \{0,1\}$.
The \textit{trace function} Tr from $\mathbb{F}_4$ onto $\mathbb{F}_2$ is defined as $\textup{Tr}(a) = a + a^2$ for $a \in \mathbb{F}_4$.
The \textit{trace} Tr$(\boldsymbol{a})$ of an $n$-dimensional vector $\boldsymbol{a} = (a_0, \dots, a_{n-1}) \in \mathbb{F}_4^n$ is the $n$-dimensional vector
$\textup{Tr}(\boldsymbol{a}) = (\textup{Tr}(a_0),\dots, \textup{Tr}(a_{n-1})) \in \mathbb{F}_2^n$.
Note that $\textup{Tr}$ is $\mathbb{F}_4$-additive and $\mathbb{F}_2$-linear,
which means that for any $x, y \in \mathbb{F}_2$ and any $\boldsymbol{a}, \boldsymbol{b} \in \mathbb{F}_4^n$, we have
$\textup{Tr}(x\boldsymbol{a}+y\boldsymbol{b}) = x\textup{Tr}(\boldsymbol{a})+y\textup{Tr}(\boldsymbol{b})$.
It is readily checked that any $n$-dimensional vector $\boldsymbol{a} \in \mathbb{F}_4^n$ can be expressed
by using $\textup{Tr}$ as
$\boldsymbol{a} = \omega^2\textup{Tr}(\boldsymbol{a}) +  \textup{Tr}(\omega\boldsymbol{a})$.
We define the trace of a column vector the same way, so that $\textup{Tr}(\boldsymbol{a}^T) = (\textup{Tr}(\boldsymbol{a}))^T$.
The $\mathbb{F}_4$-additivity and $\mathbb{F}_2$-linearity of $\textup{Tr}$
implies that for any $\boldsymbol{a} \in \mathbb{F}_4^n$ and any binary matrix $A$ with $n$ columns, we have
$\textup{Tr}(A\boldsymbol{a}^T) = A(\textup{Tr}(\boldsymbol{a}))^T$.

Let $\mathcal{C} \subseteq \mathbb{F}_4^n$ be a quaternary linear $[n,k,d]_4$ code over $\mathbb{F}_4$
and $H$ its full-rank $(n-k) \times n$ parity-check matrix.
Without loss of generality, we assume that $H$ is in standard form with the first $n-k$ columns forming the $(n-k) \times (n-k)$ identity matrix $I$, so that
\[H = \left[\begin{array}{cc}I & A\end{array}\right]\]
for some $(n-k) \times k$ matrix $A$ over $\mathbb{F}_4$.
Define a $2(n-k) \times n$ matrix $H_Q$ as
\[H_Q = \left[\begin{array}{c}
H\\
\omega H
\end{array}\right].\]
We call $H_Q$ a \textit{trace parity-check} matrix of $\mathcal{C}$.
Take the unique decomposition $H_Q =  H_Z + \omega H_X$
into a pair $H_Z$, $H_X$ of $2(n-k)\times n$ matrices over $\mathbb{F}_2$.
We call $H_Z$ and $H_X$ the $Z$-\textit{matrix} and $X$-\textit{matrix} of $H_Q$ respectively.
By assumption, the two binary components are of the form
\[
H_Z = \left[\begin{array}{cc}
I & \multirow{2}{*}{{\large $H_Z'$}}\\
0 & 
\end{array}\right]
\]
and
\[
H_X = \left[\begin{array}{cc}
0 & \multirow{2}{*}{{\large $H_X'$}}\\
I & 
\end{array}\right]
\]
for some binary $2(n-k) \times k$ matrices $H_Z'$ and $H_X'$, where $0$ is the $(n-k) \times (n-k)$ zero matrix.

For an $n$-dimensional vector $\boldsymbol{a} \in \mathbb{F}_4^n$,
we call $\textup{Tr}\left(H_Q\boldsymbol{a}^T\right)$ the \textit{trace syndrome} of $\boldsymbol{a}$.
As usual, the \textit{support} of an $m$-dimensional vector $\boldsymbol{a} = (a_0,\dots, a_{m-1})$ over a finite field is the set
$\text{supp}(\boldsymbol{a}) = \{i \ \vert\ a_i \not= 0\}$ of the coordinates at which entries are nonzero.
We use the following propositions (see the appendix for their proofs):
\begin{proposition}\label{P4}
Let $H_Z$ and $H_X$ be the $Z$-matrix and $X$-matrix of a trace parity-check matrix $H_Q$ of a linear code $\mathcal{C}$ over $\mathbb{F}_4$.
For any $\boldsymbol{a} \in \mathbb{F}_4^n$, the trace syndrome can be expressed as
$\textup{Tr}(H_Q\boldsymbol{a}^T) = H_Z\textup{Tr}(\boldsymbol{a}^T) + H_X\textup{Tr}(\omega\boldsymbol{a}^T)$.
\end{proposition}
\begin{proposition}\label{P3}
Let $\mathcal{C}$ be a linear $[n,k,d]_4$ code over $\mathbb{F}_4$ and $H_Q$ its trace parity-check matrix.
For any pair $\boldsymbol{e}, \boldsymbol{e}' \in \mathbb{F}_4^n$ of distinct $n$-dimensional vectors such that
$\vert\textup{supp}(\boldsymbol{e})\vert, \vert\textup{supp}(\boldsymbol{e}')\vert \leq \lfloor\frac{d-1}{2}\rfloor$,
their trace syndromes are distinct, that is, $\textup{Tr}(H_Q\boldsymbol{e}^T) \not= \textup{Tr}(H_Q{\boldsymbol{e}'}^T)$.
\end{proposition}

\textit{Quantum error correction.---}
Now we describe our quantum error correction scheme.

Let $\ket{0}^{\otimes 2(n-k)}_X$ be $2(n-k)$ qubits in the joint $+1$ eigenstate of $X^{\otimes 2(n-k)}$.
Without loss of generality, we assume that $\ket{0}_X = \frac{\ket{0}+\ket{1}}{\sqrt{2}}$ and that $\ket{1}_X = \frac{\ket{0}-\ket{1}}{\sqrt{2}}$,
where $\ket{0}$ and $\ket{1}$ are the computational basis.

For a unitary operator $U$ and a binary vector $\boldsymbol{a} = (a_0, \dots, a_{k-1}) \in \mathbb{F}_2^k$,
define $U^{\boldsymbol{a}}$ as the $k$-fold tensor product $O_0 \otimes \dots\otimes O_{k-1}$, where
$O_i = U$ if $a_i = 1$ and $O_i$ is the identity operator otherwise.

Our main results will be derived from the following lemma:
\begin{lemma}\label{qeclemma}
Let $\mathcal{C}$ be a linear $[n,k]_4$ code over $\mathbb{F}_4$ and $H_Q$ its trace parity-check matrix.
Let
\[\left[\begin{array}{cc}
I & \multirow{2}{*}{{\large $H_Z'$}}\\
0 & 
\end{array}\right]\]
and
\[\left[\begin{array}{cc}
0 & \multirow{2}{*}{{\large $H_X'$}}\\
I & 
\end{array}\right]\]
be the $Z$-matrix and $X$-matrix of $H_Q$ respectively.
Take an arbitrary $k$ qubit state $\ket{\psi}$.
Define unitary operator
\[Q = \sum_{\mu \in \mathbb{F}_2^{2(n-k)}}\ket{\mu}\bra{\mu}\otimes X^{\mu H_X'}Z^{\mu H_Z'}\]
on $2n-k$ qubits.
Take a pair $\boldsymbol{e}_X, \boldsymbol{e}_Z \in \mathbb{F}_2^{2n-k}$ of arbitrary $(2n-k)$-dimensional vectors.
Define ${\boldsymbol{e}_X}_l$ and ${\boldsymbol{e}_X}_r$ as
the first $2(n-k)$ and the remaining $k$ bits of ${\boldsymbol{e}_X}$ respectively so that $\boldsymbol{e}_X = ({\boldsymbol{e}_X}_l, {\boldsymbol{e}_X}_r)$.
Define similarly $\boldsymbol{e}_Z = ({{\boldsymbol{e}_Z}_l}_0, {{\boldsymbol{e}_Z}_l}_1, {\boldsymbol{e}_Z}_r)$,
where ${{\boldsymbol{e}_Z}_l}_0$, ${{\boldsymbol{e}_Z}_l}_1$, and ${\boldsymbol{e}_Z}_r$ are
the first $n-k$, the next $n-k$, and the last $k$ bits of ${\boldsymbol{e}_Z}$ respectively.
Let $\boldsymbol{e} = \omega^2({\boldsymbol{e}_Z}_{l_0}, {\boldsymbol{e}_X}_r) + ({\boldsymbol{e}_Z}_{l_1}, {\boldsymbol{e}_Z}_r)$.
Then
\begin{align*}
Q^{\dag}X^{\boldsymbol{e}_X}Z^{\boldsymbol{e}_Z}&Q\ket{0}^{\otimes 2(n-k)}_X\ket{\psi}\\
&= \ket{\textup{Tr}\left(H_Q\boldsymbol{e}^T\right) + H_Z'{H_X'}^T{{\boldsymbol{e}_X}_l}^T}_X\\
&\quad \otimes X^{{\boldsymbol{e}_X}_l H_X' + {\boldsymbol{e}_X}_r}Z^{{\boldsymbol{e}_X}_l H_Z' + {\boldsymbol{e}_Z}_r}\ket{\psi}.
\end{align*}
\end{lemma}
\Proof Let ${\boldsymbol{e}_Z}_l = ({{\boldsymbol{e}_Z}_l}_0, {{\boldsymbol{e}_Z}_l}_1)$ be the first $2(n-k)$ bits of $\boldsymbol{e}_Z$.
Then
\begin{align}\label{eq1}
Q^{\dag}&X^{\boldsymbol{e}_X}Z^{\boldsymbol{e}_Z}Q\ket{0}^{\otimes 2(n-k)}_X\ket{\psi}\notag\\
&= \ket{H_Z'{{\boldsymbol{e}_X}_r}^T + H_X'{{\boldsymbol{e}_Z}_r}^T + H_Z'{H_X'}^T{{\boldsymbol{e}_X}_l}^T + {{\boldsymbol{e}_Z}_l}^T}_X\notag\\
&\quad \otimes X^{{\boldsymbol{e}_X}_l H_X' + {\boldsymbol{e}_X}_r}Z^{{\boldsymbol{e}_X}_l H_Z' + {\boldsymbol{e}_Z}_r}\ket{\psi}
\end{align}
(see the appendix for the proof).
Note that $\textup{Tr}(\boldsymbol{e}) = ({\boldsymbol{e}_Z}_{l_0}, {\boldsymbol{e}_X}_r)$
and
$\textup{Tr}(\omega\boldsymbol{e}) = ({\boldsymbol{e}_Z}_{l_1}, {\boldsymbol{e}_Z}_r)$.
By Proposition \ref{P4}, we have
\begin{align*}
\textup{Tr}\left(H_Q\boldsymbol{e}^T\right) &= H_Z\textup{Tr}\left(\boldsymbol{e}^T\right) + H_X\textup{Tr}\left(\omega\boldsymbol{e}^T\right)\\
&= \left[\begin{array}{c}
I \\
I 
\end{array}\right]({\boldsymbol{e}_Z}_{l_0}, {\boldsymbol{e}_Z}_{l_1})^T
+
H_Z'{{\boldsymbol{e}_X}_r}^T + H_X'{{\boldsymbol{e}_Z}_r}^T\\
&= {\boldsymbol{e}_Z}_l + H_Z'{{\boldsymbol{e}_X}_r}^T + H_X'{{\boldsymbol{e}_Z}_r}^T.
\end{align*}
Plugging the above equation into Equation (\ref{eq1}) proves the assertion.
\qed

If we have $2(n-k)$ qubits that may experience phase errors but no bit errors,
then we may assume that ${\boldsymbol{e}_X}_l = 0$ in Lemma \ref{qeclemma}.
With this assumption, we are able to prove Theorem \ref{onlyZ} stated earlier.
In fact, the unitary operation $Q$ and its Hermitian conjugate serve as encoding and decoding operations
that protect arbitrary $k$ qubit quantum information from general quantum errors:
\setcounter{theorem}{0}
\begin{theorem}
If there exists a linear $[n,k,d]_4$ code over $\mathbb{F}_4$,
then there exit unitary operations that encode $k$ logical qubits into $2n-k$ physical qubits and correct up to $\lfloor\frac{d-1}{2}\rfloor$ quantum errors
under the assumption that a fixed set of $2(n-k)$ physical qubits may experience phase errors but no bit errors.
\end{theorem}\setcounter{theorem}{4}
\Proof
Let $\mathcal{C}$ be a linear $[n,k,d]_4$ code over $\mathbb{F}_4$.
We encode arbitrary $k$ qubit state $\ket{\psi}$ with $2(n-k)$ ancilla qubits $\ket{0}^{\otimes 2(n-k)}_X$
by applying $Q$ defined in Lemma \ref{qeclemma}, so that the encoding transformation is
\[\ket{\psi} \rightarrow Q\ket{0}^{\otimes 2(n-k)}_X\ket{\psi}.\]
Assume that the $2(n-k)$ ancilla qubits may experience phase errors but no bit errors while any quantum error may occur on the remaining $k$ qubits.
We regard the two binary vectors $\boldsymbol{e}_X = ({\boldsymbol{e}_X}_l, {\boldsymbol{e}_X}_r)$ and
$\boldsymbol{e}_Z = ({{\boldsymbol{e}_Z}_l}_0, {{\boldsymbol{e}_Z}_l}_1, {\boldsymbol{e}_Z}_r)$ defined in Lemma \ref{qeclemma}
as error vectors that specify the positions of discretized quantum errors,
so that the assumption on possible errors on ancilla qubits translates into the condition that ${\boldsymbol{e}_X}_l = 0$.
Our objective is to transform the state $X^{\boldsymbol{e}_X}Z^{\boldsymbol{e}_Z}Q\ket{0}^{\otimes 2(n-k)}_X\ket{\psi}$ to the original state $\ket{\psi}$
under the assumptions that $\vert \text{supp}(\boldsymbol{e}_X \cup \boldsymbol{e}_Z) \vert \leq \lfloor\frac{d-1}{2}\rfloor$
and that ${\boldsymbol{e}_X}_l = 0$.
We use the Hermitian conjugate $Q^{\dag}$ of the encoding operator $Q$ as our decoding operator.
By Lemma \ref{qeclemma}, we have
\begin{align*}
Q^{\dag}X^{\boldsymbol{e}_X}Z^{\boldsymbol{e}_Z}Q&\ket{0}^{\otimes 2(n-k)}_X\ket{\psi}\\
&= \ket{\textup{Tr}\left(H_Q\boldsymbol{e}^T\right) + H_Z'{H_X'}^T{{\boldsymbol{e}_X}_l}^T}_X\\
&\quad \otimes X^{{\boldsymbol{e}_X}_l H_X' + {\boldsymbol{e}_X}_r}Z^{{\boldsymbol{e}_X}_l H_Z' + {\boldsymbol{e}_Z}_r}\ket{\psi}\\
&= \ket{\textup{Tr}\left(H_Q\boldsymbol{e}^T\right)}_X
\otimes X^{{\boldsymbol{e}_X}_r}Z^{{\boldsymbol{e}_Z}_r}\ket{\psi}.
\end{align*}
Note that
\begin{align*}
\vert \text{supp}(\boldsymbol{e}) \vert
&= \left\vert\text{supp}\left(({\boldsymbol{e}_Z}_{l_0}, {\boldsymbol{e}_X}_r)\right)\cup\text{supp}\left(({\boldsymbol{e}_Z}_{l_1}, {\boldsymbol{e}_Z}_r)\right)\right\vert\\
&\leq \left\vert \text{supp}(\boldsymbol{e}_X) \cup \text{supp}(\boldsymbol{e}_Z) \right\vert\\
&\leq \left\lfloor\frac{d-1}{2}\right\rfloor.
\end{align*}
Thus, by Proposition \ref{P3}, measuring the ancilla qubits in the Hadamard rotated basis uniquely identifies
$\boldsymbol{e} = \omega^2({\boldsymbol{e}_Z}_{l_0}, {\boldsymbol{e}_X}_r) + ({\boldsymbol{e}_Z}_{l_1}, {\boldsymbol{e}_Z}_r)$.
Applying Pauli operators $X$ and $Z$ accordingly gives the tensor product of the original state $\ket{\psi}$ and ancilla qubits $\ket{0}^{\otimes 2(n-k)}_X$.
\qed

The above theorem can be extended to the case when the underlying classical linear code is binary:
\begin{theorem}\label{binaryonlyZ}
If there exists a linear $[n,k,d]_2$ code over $\mathbb{F}_2$,
then there exit unitary operations that encode $k$ logical qubits into $2n-k$ physical qubits and correct up to $\lfloor\frac{d-1}{2}\rfloor$ quantum errors
under the assumption that a fixed set of $2(n-k)$ physical qubits may experience phase errors but no bit errors.
\end{theorem}
\Proof
Let $\mathcal{C}$ be a linear $[n,k,d]_2$ code over $\mathbb{F}_2$ and $H$ its full-rank parity-check matrix.
Without loss of generality, we assume that $H$ is in standard form
\[H = \left[\begin{array}{cc}I & A\end{array}\right]\]
for some $(n-k) \times k$ matrix $A$ over $\mathbb{F}_2$.
Define two $2(n-k) \times k$ matrices $H_Z'$ and $H_X'$ as
\[H_Z = \left[\begin{array}{c}
A\\
0
\end{array}\right]\]
and
\[H_X = \left[\begin{array}{c}
0\\
A
\end{array}\right]\]
respectively.
We use $2(n-k)$ ancilla qubits $\ket{0}^{\otimes 2(n-k)}_X$ with encoding operator
\[Q = \sum_{\mu \in \mathbb{F}_2^{2(n-k)}}\ket{\mu}\bra{\mu}\otimes X^{\mu H_X}Z^{\mu H_Z}\]
and decoding operator $Q^{\dag}$.
Define error vectors $\boldsymbol{e}_X = ({\boldsymbol{e}_X}_l, {\boldsymbol{e}_X}_r)$ and
$\boldsymbol{e}_Z = ({{\boldsymbol{e}_Z}_l}_0, {{\boldsymbol{e}_Z}_l}_1, {\boldsymbol{e}_Z}_r)$ as in Theorem \ref{onlyZ}.
Let $\boldsymbol{e}_0 = ({{\boldsymbol{e}_Z}_l}_0, {\boldsymbol{e}_X}_r)$, $\boldsymbol{e}_1 = ({{\boldsymbol{e}_Z}_l}_1, {\boldsymbol{e}_Z}_r)$,
and ${\boldsymbol{e}_Z}_l = ({{\boldsymbol{e}_Z}_l}_0, {{\boldsymbol{e}_Z}_l}_1)$ respectively.
Assume that ${\boldsymbol{e}_X}_l = 0$.
It is routine to show that for arbitrary $k$ qubit state $\ket{\psi}$,
\begin{align*}
Q^{\dag}&X^{\boldsymbol{e}_X}Z^{\boldsymbol{e}_Z}Q\ket{0}^{\otimes 2(n-k)}_X\ket{\psi}\\
&= \ket{H_Z{{\boldsymbol{e}_X}_r}^T + H_X{{\boldsymbol{e}_Z}_r}^T + {{\boldsymbol{e}_Z}_l}^T}_X
\otimes X^{{\boldsymbol{e}_X}_r}Z^{{\boldsymbol{e}_Z}_r}\ket{\psi}\\
&= \ket{\left(H{\boldsymbol{e}_0}^T, H{\boldsymbol{e}_1}^T\right)}_X
\otimes X^{{\boldsymbol{e}_X}_r}Z^{{\boldsymbol{e}_Z}_r}\ket{\psi}.
\end{align*}
If $\left\vert \text{supp}(\boldsymbol{e}_X) \cup \text{supp}(\boldsymbol{e}_Z) \right\vert \leq \left\lfloor\frac{d-1}{2}\right\rfloor$,
then $\left\vert \text{supp}(\boldsymbol{e}_0)\right\vert, \left\vert \text{supp}(\boldsymbol{e}_1)\right\vert \leq  \left\lfloor\frac{d-1}{2}\right\rfloor$.
Thus, as in standard syndrome decoding for classical linear codes,
measuring the ancilla qubits uniquely identifies the quantum errors that occurred on the $2n-k$ qubits.
\qed

Trivially, if we assume that the $2(n-k)$ ancilla qubits in the above two theorems are free from quantum errors,
our quantum error correction scheme is an entanglement-assisted quantum error-correcting code
that encodes $k$ logical qubits into $k$ physical qubits with the help of $2(n-k)$ ebits.
By following the notation in \cite{BDH}, we have the following corollary:
\begin{corollary}\label{eaqecc}
If there exists a linear $[n,k,d]_2$ code over $\mathbb{F}_2$ or a linear $[n,k,d]_4$ code over $\mathbb{F}_4$,
then there exists an entanglement-assisted quantum $[[k,k,d';2(n-k)]]$ error-correcting code, where $d' \geq d$.
\end{corollary}

The quantum Singleton bound gives an upper bound on the error correction capability of a quantum error-correcting code:
\begin{theorem}[\cite{KL,BDH}]\label{Singleton}
Let $n_e$ and $d_e$ be positive integers such that $n_e \geq 2(d_e-1)$.
For any entanglement-assisted quantum $[[n_e,k_e,d_e;c]]$ code, $k_e - c \leq n_e-2d_e+2$.
\end{theorem}

Our quantum error correction scheme can provide the best possible entanglement-assisted error correction capability
if the underlying classical codes have the largest possible minimum distances:
\begin{theorem}
If there exists a nontrivial \textup{MDS} code of length $n$ and dimension $k$ over $\mathbb{F}_4$,
then there exists an entanglement-assisted quantum error-correcting code of length $k$, dimension $k$, and distance $n-k+1$
that saturates the quantum Singleton bound.
\end{theorem}
\Proof
The minimum distance of an MDS code of length $n$ and dimension $k$ is $n-k+1$.
By Corollary \ref{eaqecc}, an MDS code gives an entanglement-assisted quantum
$[[n_e,k_e,d_e;c]]$ error-correcting code with $n_e = k_e = k$, $d_e \geq n-k+1$, and $c = 2(n-k)$.
Thus, we have
\begin{align*}
n_e-2d_e+2 &\leq k - 2(n-k+1) + 2\\
&= 3k-2n\\
&= k_e-c,
\end{align*}
saturating the quantum Singleton bound.
\qed

It is straightforward to see that classical linear codes can be imported the same way for when ancilla qubits suffer only bit errors
by Hadamard rotating the ancilla qubits and corresponding outer products in the encoding and decoding operators.

\textit{Concluding remarks.---}
Because the dominating type of quantum error will likely be difficult to completely suppress through hardware in actual implementations of quantum devices,
it is essential for software solutions to assume their occurrences.
Our theoretical analysis shows that this assumption can be good enough
to correct general quantum errors through classical error-correcting codes.
This makes the requirements on hardware less demanding
because we only have to make part of the system more reliable than the rest.
Another notable property of our scheme is that the number of required less noisy qubits only depends on the length and dimension of the underlying classical code.
In fact, a higher dimension leads to a fewer less noisy qubits.

An interesting question we did not address is whether or how much efficient encoding and decoding procedures for classical linear codes
help develop equally efficient procedures for our quantum error-correcting codes.
While this is not a trivial question, looking at the simple correspondence between quantum and classical error-correcting codes
and the recent development on efficiently decodable quantum error-correcting codes \cite{WSBW,HYH2},
research in this direction appears promising.
We hope that our finding facilitates the development towards large-scale quantum computation and quantum communication.

\begin{acknowledgments}
Y.F. acknowledges support from JSPS
and thanks the anonymous referee for their careful reading and valuable comments.
\end{acknowledgments}

\newpage
\onecolumngrid
\appendix
\section*{Proofs of Propositions 2 and 3}
This section gives the proof of Propositions 2 and 3. For convenience, we reiterate the statements of the propositions here.
\setcounter{theorem}{1}
\begin{proposition}
Let $H_Z$ and $H_X$ be the $Z$-matrix and $X$-matrix of a trace parity-check matrix $H_Q$ of a linear code $\mathcal{C}$ over $\mathbb{F}_4$.
For any $\boldsymbol{a} \in \mathbb{F}_4^n$, the trace syndrome can be expressed as
$\textup{Tr}(H_Q\boldsymbol{a}^T) = H_Z\textup{Tr}(\boldsymbol{a}^T) + H_X\textup{Tr}(\omega\boldsymbol{a}^T)$.
\end{proposition}
\Proof Let $\mathcal{C}$ be a linear code over $\mathbb{F}_4$ and $H_Q$ its trace parity-check matrix.
Assume that $H_Q = H_Z + \omega H_X$, where $H_Z$ and $H_X$ are the $Z$-matrix and $X$-matrix of $H$ respectively.
Recall that any vector $\boldsymbol{a}$ over $\mathbb{F}_4$ can be expressed as
$\boldsymbol{a} = \omega^2\textup{Tr}(\boldsymbol{a}) +  \textup{Tr}(\omega\boldsymbol{a})$.
Because the trace function $\textup{Tr}$ is $\mathbb{F}_4$-additive and $\mathbb{F}_2$-linear, we have
\begin{eqnarray*}
\textup{Tr}(H_Q\boldsymbol{a}^T) &=& \textup{Tr}\left((H_Z+\omega H_X)(\omega^2\textup{Tr}(\boldsymbol{a})+\textup{Tr}(\omega\boldsymbol{a}))^T\right)\\
&=& H_Z\textup{Tr}\left(\omega^2\textup{Tr}(\boldsymbol{a}^T)+\textup{Tr}(\omega\boldsymbol{a}^T)\right)
+ H_X\textup{Tr}\left(\textup{Tr}(\boldsymbol{a}^T)+\omega\textup{Tr}(\omega\boldsymbol{a}^T)\right)\\
&=& H_Z\textup{Tr}(\boldsymbol{a}^T) + H_X\textup{Tr}(\omega\boldsymbol{a}^T).
\end{eqnarray*}
The proof is complete.
\qed

\begin{proposition}
Let $\mathcal{C}$ be a linear $[n,k,d]_4$ code over $\mathbb{F}_4$ and $H_Q$ its trace parity-check matrix.
For any pair of distinct $n$-dimensional vectors $\boldsymbol{e}, \boldsymbol{e}' \in \mathbb{F}_4^n$ such that
$\vert\textup{supp}(\boldsymbol{e})\vert, \vert\textup{supp}(\boldsymbol{e}')\vert \leq \lfloor\frac{d-1}{2}\rfloor$,
their trace syndromes are distinct, that is, $\textup{Tr}(H_Q\boldsymbol{e}^T) \not= \textup{Tr}(H_Q{\boldsymbol{e}'}^T)$.
\end{proposition}
\Proof Let $\mathcal{C}$ be a linear $[n,k,d]_4$ code over $\mathbb{F}_4$
and $H_Q$ its trace parity-check matrix composed of parity-check matrix $H$ of $\mathcal{C}$ and $\omega H$.
Because the rows of $H_Q$ consist of those of $H$ and $\omega H$,
the condition $\textup{Tr}(H_Q\boldsymbol{c}^T) = 0$ holds if and only if
the two conditions that $\textup{Tr}(H\boldsymbol{c}^T) = 0$ and that $\textup{Tr}(\omega H\boldsymbol{c}^T) = 0$ simultaneously hold.
Because $\textup{Tr}(0) = \textup{Tr}(1) = 0$ and $\textup{Tr}(\omega) = \textup{Tr}(\omega^2) = 1$,
the condition $\textup{Tr}(H\boldsymbol{c}^T) = 0$ implies that $(H\boldsymbol{c}^T)$ is a column vector over $\mathbb{F}_2 = \{0,1\}$.
If $\text{supp}(H\boldsymbol{c}^T) \not=0$, then $\textup{Tr}(\omega H\boldsymbol{c}^T) \not= 0$.
Thus, the conditions $\textup{Tr}(H\boldsymbol{c}^T) = 0$ and $\textup{Tr}(\omega H\boldsymbol{c}^T) = 0$ simultaneously hold
if and only if $H\boldsymbol{c}^T = 0$, which means that $\boldsymbol{c}$ is a codeword of $\mathcal{C}$.
Take a pair $\boldsymbol{e}, \boldsymbol{e}' \in \mathbb{F}_4^n$ of distinct $n$-dimensional vectors such that
$\vert\textup{supp}(\boldsymbol{e})\vert, \vert\textup{supp}(\boldsymbol{e}')\vert \leq \lfloor\frac{d-1}{2}\rfloor$.
Suppose to the contrary that $\textup{Tr}(H_Q\boldsymbol{e}) = \textup{Tr}(H_Q\boldsymbol{e}')$.
Then because the trace function $\textup{Tr}$ is $\mathbb{F}_4$-additive and $\mathbb{F}_2$-linear,
we have
\begin{eqnarray*}
0 &=& \textup{Tr}(H_Q\boldsymbol{e}^T) + \textup{Tr}(H_Q{\boldsymbol{e}'}^T)\\
&=& \textup{Tr}(H_Q(\boldsymbol{e}+\boldsymbol{e}')^T),
\end{eqnarray*}
which implies that $\boldsymbol{e}+\boldsymbol{e}' \in \mathcal{C}$.
However, because
\begin{eqnarray*}
\vert\textup{supp}(\boldsymbol{e}+\boldsymbol{e}')\vert &\leq&
\vert\textup{supp}(\boldsymbol{e})\vert + \vert\textup{supp}(\boldsymbol{e}')\vert\\
&\leq& 2\left\lfloor\frac{d-1}{2}\right\rfloor\\
&\leq& d-1,
\end{eqnarray*}
this is a contradiction. This completes the proof.
\qed

\section*{Derivation of the equation in the proof of Lemma 4}
Here we derive Equation (1) in the proof of Lemma 4.
In what follows, we do not explicitly write the global phase factor $e^{i\theta}$.

\noindent\textbf{Derivation of Equation (1).}\quad
\begin{align*}
Q^{\dag}X^{\boldsymbol{e}_X}&Z^{\boldsymbol{e}_Z}Q\ket{0}^{\otimes 2(n-k)}_X\ket{\psi}\\
&= Q^{\dag}X^{\boldsymbol{e}_X}Z^{\boldsymbol{e}_Z}\left(\sum_{\mu \in \mathbb{F}_2^{2(n-k)}}\ket{\mu}\bra{\mu}
\otimes X^{\mu H_X'}Z^{\mu H_Z'}\right)\ket{0}^{\otimes 2(n-k)}_X\ket{\psi}\notag\\
&= 2^{k-n}Q^{\dag}X^{\boldsymbol{e}_X}Z^{\boldsymbol{e}_Z}\sum_{\mu \in \mathbb{F}_2^{2(n-k)}}\ket{\mu}\otimes X^{\mu H_X'}Z^{\mu H_Z'}\ket{\psi}\\
&= 2^{k-n}Q^{\dag}\sum_{\mu \in \mathbb{F}_2^{2(n-k)}}X^{{\boldsymbol{e}_X}_l}Z^{{\boldsymbol{e}_Z}_l}\ket{\mu}
\otimes (-1)^{\mu H_Z'{{\boldsymbol{e}_X}_r}^T + \mu H_X'{{\boldsymbol{e}_Z}_r}^T}
X^{\mu H_X'}Z^{\mu H_Z'}X^{{\boldsymbol{e}_X}_r}Z^{{\boldsymbol{e}_Z}_r}\ket{\psi}\\
&= 2^{k-n}\left(\sum_{\lambda \in \mathbb{F}_2^{2(n-k)}}\ket{\lambda}\bra{\lambda}\otimes (X^{\lambda H_X'}Z^{\lambda H_Z'})^{\dag}\right)
\sum_{\mu \in \mathbb{F}_2^{2(n-k)}}X^{{\boldsymbol{e}_X}_l}Z^{{\boldsymbol{e}_Z}_l}\ket{\mu}
\otimes (-1)^{\mu H_Z'{{\boldsymbol{e}_X}_r}^T + \mu H_X'{{\boldsymbol{e}_Z}_r}^T}
X^{\mu H_X'}Z^{\mu H_Z'}X^{{\boldsymbol{e}_X}_r}Z^{{\boldsymbol{e}_Z}_r}\ket{\psi}\\
&= 2^{k-n}\sum_{\mu \in \mathbb{F}_2^{2(n-k)}}(-1)^{\mu H_Z'{{\boldsymbol{e}_X}_r}^T+\mu H_X'{{\boldsymbol{e}_Z}_r}^T}
X^{{\boldsymbol{e}_X}_l}Z^{{\boldsymbol{e}_Z}_l}\ket{\mu}
\otimes (X^{(\mu+{\boldsymbol{e}_X}_l)H_X'}Z^{(\mu+{\boldsymbol{e}_X}_l)H_Z'})^{\dag}
X^{\mu H_X'}Z^{\mu H_Z'}X^{{\boldsymbol{e}_X}_r}Z^{{\boldsymbol{e}_Z}_r}\ket{\psi}\\
&= 2^{k-n}\sum_{\mu \in \mathbb{F}_2^{2(n-k)}}(-1)^{\mu H_Z'{{\boldsymbol{e}_X}_r}^T+\mu H_X'{{\boldsymbol{e}_Z}_r}^T}
X^{{\boldsymbol{e}_X}_l}Z^{{\boldsymbol{e}_Z}_l}\ket{\mu}
\otimes Z^{\mu H_Z'+{\boldsymbol{e}_X}_l H_Z'}X^{\mu H_X'+{\boldsymbol{e}_X}_l H_X'}
X^{\mu H_X'}Z^{\mu H_Z'}X^{{\boldsymbol{e}_X}_r}Z^{{\boldsymbol{e}_Z}_r}\ket{\psi}\notag\\
&=  2^{k-n}\sum_{\mu \in \mathbb{F}_2^{2(n-k)}}(-1)^{\mu H_Z'{{\boldsymbol{e}_X}_r}^T+\mu H_X'{{\boldsymbol{e}_Z}_r}^T}
X^{{\boldsymbol{e}_X}_l}Z^{{\boldsymbol{e}_Z}_l}\ket{\mu}
\otimes Z^{\mu H_Z'+{\boldsymbol{e}_X}_l H_Z'}X^{{\boldsymbol{e}_X}_l H_X'}
Z^{\mu H_Z'}X^{{\boldsymbol{e}_X}_r}Z^{{\boldsymbol{e}_Z}_r}\ket{\psi}\\
&= 2^{k-n}\sum_{\mu \in \mathbb{F}_2^{2(n-k)}}(-1)^{\mu H_Z'{{\boldsymbol{e}_X}_r}^T+\mu H_X'{{\boldsymbol{e}_Z}_r}^T+\mu H_Z' ({\boldsymbol{e}_X}_l H_X')^T}
X^{{\boldsymbol{e}_X}_l}Z^{{\boldsymbol{e}_Z}_l}\ket{\mu}
\otimes Z^{\mu H_Z'+{\boldsymbol{e}_X}_l H_Z'}
Z^{\mu H_Z'}X^{{\boldsymbol{e}_X}_l H_X'}X^{{\boldsymbol{e}_X}_r}Z^{{\boldsymbol{e}_Z}_r}\ket{\psi}\\
&= 2^{k-n}\sum_{\mu \in \mathbb{F}_2^{2(n-k)}}(-1)^{\mu H_Z'{{\boldsymbol{e}_X}_r}^T+\mu H_X'{{\boldsymbol{e}_Z}_r}^T + \mu H_Z' ({\boldsymbol{e}_X}_l H_X')^T}
X^{{\boldsymbol{e}_X}_l}Z^{{\boldsymbol{e}_Z}_l}\ket{\mu}
\otimes Z^{{\boldsymbol{e}_X}_l H_Z'}
X^{{\boldsymbol{e}_X}_l H_X'}X^{{\boldsymbol{e}_X}_r}Z^{{\boldsymbol{e}_Z}_r}\ket{\psi}\\
&= 2^{k-n}\sum_{\mu \in \mathbb{F}_2^{2(n-k)}}(-1)^{\mu H_Z'{{\boldsymbol{e}_X}_r}^T+\mu H_X'{{\boldsymbol{e}_Z}_r}^T + \mu H_Z'({\boldsymbol{e}_X}_l H_X')^T + \mu {{\boldsymbol{e}_Z}_l}^T}
X^{{\boldsymbol{e}_X}_l}\ket{\mu}
\otimes Z^{{\boldsymbol{e}_X}_l H_Z'}
X^{{\boldsymbol{e}_X}_l H_X'}X^{{\boldsymbol{e}_X}_r}Z^{{\boldsymbol{e}_Z}_r}\ket{\psi}\\
&= X^{{\boldsymbol{e}_X}_l}\ket{H_Z'{{\boldsymbol{e}_X}_r}^T + H_X'{{\boldsymbol{e}_Z}_r}^T + H_Z'({\boldsymbol{e}_X}_l H_X')^T + {{\boldsymbol{e}_Z}_l}^T}_X
\otimes Z^{{\boldsymbol{e}_X}_l H_Z'}
X^{{\boldsymbol{e}_X}_l H_X'}X^{{\boldsymbol{e}_X}_r}Z^{{\boldsymbol{e}_Z}_r}\ket{\psi}\\
&= \ket{H_Z'{{\boldsymbol{e}_X}_r}^T + H_X'{{\boldsymbol{e}_Z}_r}^T + H_Z'{H_X'}^T{{\boldsymbol{e}_X}_l}^T + {{\boldsymbol{e}_Z}_l}^T}_X
\otimes X^{{\boldsymbol{e}_X}_l H_X' + {\boldsymbol{e}_X}_r}Z^{{\boldsymbol{e}_X}_l H_Z' + {\boldsymbol{e}_Z}_r}\ket{\psi}.
\end{align*}

\end{document}